\documentclass{llncs}
\usepackage{amsmath}
\usepackage{amsfonts}
\usepackage{amssymb}
\usepackage[utf8]{inputenc}
\usepackage{xcolor}
\usepackage{algorithm}
\usepackage{algpseudocode}
\usepackage{array, multirow}
\usepackage{subfig}
\usepackage{tikz}
\usepackage{pgfplots}

\newcommand{\Z}[1]{\mathbb{Z}/(#1\mathbb{Z})}

\newcommand{\ket}[1]{\left|#1\right>}

\newcommand{\khi}[1]{\chi\left(#1\right)}
\newcommand{\s}{\raisebox{0pt}[1.2em]{}}
\newcommand{\sline}{\hline\s{}}

\newcommand{\ecversion}[1]{#1}

\title{Improved Low-qubit Hidden Shift Algorithms}

\ecversion{
\author{Xavier Bonnetain}
\institute{Sorbonne Université, Collège Doctoral, F-75005 Paris, France \and Inria, France}
}

\begin{document}
\maketitle
\begin{abstract}
 Hidden shift problems are relevant to assess the quantum security of various cryptographic constructs. Multiple quantum subexponential time algorithms have been proposed. In this paper, we propose some improvements on a polynomial quantum memory algorithm proposed by Childs, Jao and Soukharev in 2010. We use subset-sum algorithms to significantly reduce its complexity.
 We also propose new tradeoffs between quantum queries, classical time and classical memory to solve this problem.
\end{abstract}

\section{Introduction}
The hidden shift problem can be stated as follows:

{\em Let $f$, $g$ be two injective functions, $\mathbb{G}$ a group. Given the promise that there exists $s \in \mathbb{G}$ such that,
for all $x$, $f(x) = g(xs)$, retrieve $s$.}

This is a generalization of the the hidden subgroup problem, which corresponds to the case $f = g$, and is efficiently solved by Shor's algorithm~\cite{DBLP:journals/siamcomp/Shor97} in the abelian case.

As the hidden subgroup problem, the hidden shift problem is of interest for cryptography. Notably, the security of multiple symmetric primitives~\cite{conf/eurocrypt/AlagicR17,DBLP:conf/asiacrypt/BonnetainN18} and the security of some post-quantum isogeny-based asymmetric schemes~\cite{cryptoeprint:2018:383,DBLP:journals/jmc/ChildsJS14,DBLP:journals/iacr/FeoG18,DBLP:conf/asiacrypt/FeoKS18} depends on its hardness.

It is also interesting for quantum computing, as solving this problem requires exponential classical time, but depending on the group structure, can be solved quantumly in either polynomial, sub-exponential or exponential time.

The first subexponential quantum algorithm for the hidden shift problem has been proposed by Kuperberg in~\cite{DBLP:journals/siamcomp/Kuperberg05}, where he proposed multiple algorithms in $2^{O(\sqrt{n})}$ quantum time, memory and query. A polynomial quantum memory variant has been proposed by Regev in~\cite{quant-ph/0406151v1}, with a time cost in $2^{O(\sqrt{n\log_2(n)})}$. This latter algorithm has been generalized and more precisely studied by Childs, Jao and Soukharev in~\cite{DBLP:journals/jmc/ChildsJS14}, where they prove a time cost in $2^{(\sqrt{2}+o(1))\sqrt{n\log_2(n)}}$ for all abelian groups. In 2013, Kuperberg proposed a generalisation of his first algorithm and Regev's variant~\cite{DBLP:conf/tqc/Kuperberg13} with a heuristic cost estimate of $\widetilde{O}\left(2^{\sqrt{2n}}\right)$. 

Asymptotic cost estimates provide little information on the concrete cost of the algorithm. Concrete cost estimates for groups of order a power of two for Kuperberg's original algorithm have been done in~\cite{DBLP:conf/asiacrypt/BonnetainN18}, obtaining a cost of around $2^{1.8\sqrt{n}}$ for the cyclic case. The used algorithm has later been generalised to arbitrary cyclic groups in~\cite{DBLP:journals/iacr/BonnetainS18}, for a cost of around $12\times 2^{1.8\sqrt{n}}$.

In this paper, we will focus on the quantum algorithm of Childs, Jao and Soukharev~\cite{DBLP:journals/jmc/ChildsJS14} and show how it can be improved.

\subsection{Notations}
We note \[L_n(\alpha, c) = 2^{\left((c+o(1))n^\alpha\log_2(n)^{1-\alpha}\right)}.\]

We will focus here on the cyclic group $\Z{N}$, and denote $n = \log_2(N)$. However, the algorithms can also be applied to arbitrary abelian groups, using the same approach as in~\cite{DBLP:journals/jmc/ChildsJS14}.
We denote $L_n(1/2,c)$ as $L(c)$. We note $\cdot$ the inner product of vectors.

We denote $\chi(x) = \exp\left(2i\pi x\right)$ in the qubit phases.
\subsection{Results}
In this paper, we improve the quantum hidden shift algorithm of~\cite{DBLP:journals/jmc/ChildsJS14}. We use subset-sum algorithms to lower the exponent of the complexity, and propose different tradeoffs between quantum and classical cost. In particular, with $c$ the exponent to solve a subset-sum instance, we solve the hidden shift problem in $L(\sqrt{c})$ classical time and quantum queries, and if
we want a quadratic gap between the classical time cost and the quantum query cost, we can solve it in $L(\sqrt{c/3})$ quantum queries and $L(2\sqrt{c/3})$ classical time.

The results are summarized in Table~\ref{cost:c}. The quantum time is roughly the number of quantum queries, as the only non-trivial quantum operation is the oracle evaluation.
\begin{table}
\centering{}
 \begin{tabular}{|c|c|c|c|c|}
\sline
 Quantum query & Classical time & Classical memory & Subset-sum & Source\\
 \sline
 $L\left(1/\sqrt{2}\right)$ & $L\left(\sqrt{2}\right)$ & $\widetilde{O}(1)$ & exhaustive search &\cite{DBLP:journals/jmc/ChildsJS14}\\
 \sline
 $L\left(1\right)$ & $L\left(1\right)$ & $\widetilde{O}(1)$ & exhaustive search &Section~\ref{minclass}\\
 \sline
 $L\left(1/\sqrt3\right)$ & $L\left(2/\sqrt{3}\right)$ & $\widetilde{O}(1)$ & exhaustive search &Section~\ref{quad}\\
 \sline
 $L(0.539)$ & $L(0.539)$ & $L(0.539)$ & \cite{DBLP:conf/eurocrypt/BeckerCJ11} &Section~\ref{minclass}\\
 \sline
 $L(0.312)$ & $L(0.623)$ & $L(0.312)$ & \cite{DBLP:conf/eurocrypt/BeckerCJ11} &Section~\ref{quad}\\
 \sline
 $L(0.849)$ & $L(0.849)$ & $\widetilde{O}(1)$ & \cite{DBLP:conf/eurocrypt/BeckerCJ11}, poly.memory &Section~\ref{minclass}\\
 \sline
 $L(0.490)$ & $L(0.980)$ & $\widetilde{O}(1)$ & \cite{DBLP:conf/eurocrypt/BeckerCJ11}, poly.memory &Section~\ref{quad}\\
 \sline
 $O(n^2)$ & $\widetilde{O}\left(2^{0.291n}\right)$ & $\widetilde{O}\left(2^{0.291n}\right)$ & \cite{DBLP:conf/eurocrypt/BeckerCJ11} &Section~\ref{minq}\\
 \hline
\end{tabular}
\vspace{0.2cm}
\caption{Hidden shift algorithm cost}\label{cost:c}
\end{table}

\section{Hidden Shift Algorithms}
Hidden shift algorithms are in two steps: the first one uses the oracles to produce some random qubits with a specific structure (the elements), which are then refined by some combination routines until we manage to extract the value of the hidden shift.

\subsection{Element Generation}
Given a quantum oracle access to $f$ and $g$, one can compute
\[ \sum_x \ket0\ket{x}\ket{f(x)} + \ket1\ket{x}\ket{g(x)}. \]
After a measurement of the last register, one obtains
\[ \ket0\ket{x_0}+\ket1\ket{x_0+s} \]
for a given unknown $x_0$. Now, one can apply a Quantum Fourier Transform on the second register, to obtain
\[ \sum_{\ell} \ket0\chi\left(\frac{x_0\ell}N\right)\ket{\ell} + \ket1\chi\left(\frac{(x_0+s)\ell}N\right)\ket{\ell}. \]
Finally, one can perform a measurement on the second register, and obtain $\ell$ and
\[ \ket{\psi_{\ell}} = \ket0 + \khi{\frac{s\ell}N}\ket1 .\]

\subsection{Interesting Elements}
If one manages to obtain $\ket{\psi_1},\dots,\ket{\psi_{2^j}},\ket{\psi_{2^n}}$, then $s$ can be retrieved with a Quantum Fourier Transform, as \[\bigotimes_{j = 0}^ n \ket{\psi_{2^j}} = \sum_{\ell = 0}^{2^n} \khi{\frac{s\ell}N}\ket{\ell}.\]
Hence, applying an inverse Quantum Fourier Transform allows to retrieve $s$.

It is to be noted that, for groups of odd order, we only need to be able to construct $\ket{\psi_1}$: the value $2^j$ can be obtained by multiplying all the labels by $2^{-j}$ and constructing 1 from the new labels.

The situation is slightly easier if $ N = 2^n$. In that case, $\ket{\psi_{2^{n-1}}} = \ket0 + (-1)^s\ket1$ directly gives $s\mod 2$. Likewise, as noted in~\cite{DBLP:conf/asiacrypt/BonnetainN18}, both $\ket{\psi_{2^{n-2}}}$ and $\ket{\psi_{3\times2^{n-2}}}$ allows to obtain the second bit of $s$ if $s\mod 2$ is known, and so on, knowing the lower bits of $s$ allows to extract the next bit from $\ket{\psi_{(2\alpha+1)2^j}}$.

\subsection{Combination routines}
We will use the combination routines of~\cite{DBLP:journals/jmc/ChildsJS14} to obtain the labels we are looking for, that is, either $\ket{\psi_1}$ or $\ket{\psi_{2^{n-1}}}$.

The idea is to take a certain amount of elements ($k$), and use them to produce one better element. Recall that the elements are of the form $\ket0 + \exp\left(2i\pi\frac{s\ell_i}N\right)\ket1$. If we tensor them,
we obtain \[\bigotimes_{i}\ket{\psi_{\ell_i}} = \sum_{j \in \{0,1\}^k} \khi{\frac{ j \cdot (\ell_1,\dots,\ell_k)}{N}}\ket{j}.\]

Now, the objective of the combination routine is to perform a partial measurement on $\ket{j}$, in order to ensure that the remaining $j$ have an interesting phase difference. After that, we only need to
know the corresponding $j$, project the state on a pair, and relabel the two values of the pair to 0 and 1 in order to obtain a better element.

Algorithm~\ref{alg:reg-comb} computes the function $\ket{j}\ket{0} \mapsto \ket{j}\ket{j \cdot (\ell_1,\dots,\ell_k) \mod 2^r}$ and measures the second register. By definition, the remaining $j$ have a phase identical modulo $2^r$.
Hence, the output label will be a multiple of $2^r$. This approach can be iterated, in order to obtain multiples of increasingly big powers of 2, and allows to reach $\ket{\psi_{2^{n-1}}}$.

Another approach to reach a specific element is to compute $\ket{\lfloor j \cdot (\ell_1,\dots,\ell_k)/M\rfloor}$. This will produce elements with a close phase (their difference will be smaller than $M$).
As before, this approach can be iterated to obtain smaller and smaller labels, until we reach $\ket{\psi_{1}}$. This is done in Algorithm~\ref{alg:cjs-comb}.

\begin{algorithm}
\begin{algorithmic}[1]
  \Statex \textbf{Input:} $(\ket{\psi_{\ell_1}},\dots, \ket{\psi_{\ell_k}}) : \forall i, 2^a|\ell_i$, $r$
  \Statex \textbf{Output:} $\ket{\psi_{\ell'}}$, $2^{r+a}|\ell'$
  \State Tensor $\bigotimes_{i}\ket{\psi_{\ell_i}} = \sum_{j \in \{0,1\}^k} \khi{\frac{ j \cdot (\ell_1,\dots,\ell_k)}{N}}\ket{j}$
  \State Add an ancilla register, apply $\ket{x}\ket{0} \mapsto \ket{x}\ket{x \cdot (\ell_1,\dots,\ell_k) \mod 2^r}$
  \State Measure the ancilla register, leaving with $$V\text{ and } \sum_{j\cdot (\ell_1,\dots,\ell_k) \mod 2^r= V } \khi{\frac{ j \cdot (\ell_1,\dots,\ell_k)}{N}}\ket{j}$$
  \State Compute the corresponding $j$\label{step:ss}
  \State Pair them, project to a pair $(j_1,j_2)$.\label{step:proj}
  \Statex The register is now $\khi{\frac{ j_1\cdot (\ell_1,\dots,\ell_k)}{N}}\ket{j_1}+\khi{\frac{ j_2 \cdot (\ell_1,\dots,\ell_k)}{N}}\ket{j_2}$
  \State Map $\ket{j_1}$ to $\ket0$, $\ket{j_2}$ to $\ket1$
  \State Return $\ket0 + \khi{\frac{(j_2-j_1)\cdot (\ell_1,\dots,\ell_k)}N}\ket1$
\end{algorithmic}
\caption{Combination routine, for powers of 2}\label{alg:reg-comb}
\end{algorithm}

\begin{algorithm}
\begin{algorithmic}[1]
  \Statex \textbf{Input:} $(\ket{\psi_{\ell_1}},\dots, \ket{\psi_{\ell_k}}), (\ell_1,\dots,\ell_k) \in [0;B)^k$, $r$
  \Statex \textbf{Output:} $\ket{\psi_{\ell'}}$, $\ell' < \sum_j \ell_j/2^{r}$
  \State Tensor $\bigotimes_{i}\ket{\psi_{\ell_i}} = \sum_{j \in \{0,1\}^k} \khi{\frac{ j \cdot (\ell_1,\dots,\ell_k)}{N}}\ket{j}$
  \State Add an ancilla register, apply $\ket{x}\ket{0} \mapsto \ket{x}\ket{\lfloor x \cdot (\ell_1,\dots,\ell_k)2^{r-1}/B\rfloor}$
  \State Measure the ancilla register, leaving with $$V\text{ and } \sum_{\lfloor j \cdot (\ell_1,\dots,\ell_k)2^{r-1}/B\rfloor = V} \khi{\frac{ j \cdot (\ell_1,\dots,\ell_k)}{N}}\ket{j}$$
  \State Compute the corresponding $j$
  \State Pair them, project to a pair $(j_1,j_2)$.\label{step:proj}
  \Statex The register is now $\khi{\frac{ j_1\cdot (\ell_1,\dots,\ell_k)}{N}}\ket{j_1}+\khi{\frac{ j_2 \cdot (\ell_1,\dots,\ell_k)}{N}}\ket{j_2}$
  \State Map $\ket{j_1}$ to $\ket0$, $\ket{j_2}$ to $\ket1$
  \State Return $\ket0 + \khi{\frac{(j_2-j_1)\cdot (\ell_1,\dots,\ell_k)}N}\ket1$
\end{algorithmic}
\caption{Combination routine, for smaller labels}\label{alg:cjs-comb}
\end{algorithm}

Both algorithms are used in~\cite{DBLP:journals/jmc/ChildsJS14}, the former to tackle cyclic groups of order a power of 2, the latter for cyclic groups of odd order.

\begin{algorithm}
\begin{algorithmic}[1]
  \Statex Input: $\sum_{x\in J}\phi(x)\ket{x}$, $(j_1, j_2) \subset J$.
  \Statex Output: $\phi(j_1)\ket{j_1} + \phi(j_2)\ket{j_2}$ or $\sum_{x\in J\setminus\{j_1,j_2\}}\phi(x)\ket{x}$
  \State Add an ancilla qubit: $\sum_{x\in J}\phi(x)\ket{x}\ket0$
  \State Apply the operator $\ket{x}\ket{0}\mapsto \ket{x}\ket{x = j_1 \vee x = j_2 }$
  \State Measure the ancilla qubit
 \end{algorithmic}
\caption{Projection routine}\label{alg:proj}
\end{algorithm}
Algorithm~\ref{alg:proj} projects on $\langle\ket{j_1},\ket{j_2}\rangle$ with probability $2/|J|$, and otherwise projects to the supplementary vector space.

\subsubsection{Finding pairs}
Finding the pairs for Algorithm~\ref{alg:reg-comb} (Step~\ref{step:ss}) consists in finding the solutions of the equation $x \cdot (\ell_1,\dots,\ell_k) \mod 2^r = V$. This is addressed by brute-force in~\cite{DBLP:journals/jmc/ChildsJS14}, for a cost of $2^k$.
For the complexity analysis, we consider that this step costs $\widetilde{O}(2^{ck})$, the brute-force case being the case $c = 1$. This brute-force approach can also be applied to Step~\ref{step:ss} of Algorithm~\ref{alg:cjs-comb}.

\subsubsection{Complexity}
Algorithm~\ref{alg:reg-comb} succeeds if, at step \ref{step:proj}, we manage to project on a pair $(j_1,j_2)$. Indeed, as they are both preimages of $V$, we have $$\ell' = (j_2-j_1)\cdot(\ell_1,\dots,\ell_k) = 0 \mod 2^m.$$
In order to achieve this, we must:
\begin{enumerate}
 \item Have at least two distinct solutions of the equation,
 \item Manage to find the solutions,
 \item Successfully project onto a pair.
\end{enumerate}
The first point requires us to have $r < k$. In \cite{quant-ph/0406151v1}, $ k = r+4 $ is used, while \cite{DBLP:journals/jmc/ChildsJS14} uses $k = r+1$. In both cases, we have a fixed probability to have at least 2 solutions: there are $2^r$ images, hence at most $2^r$ subsets have a sum for which there is a unique preimage, hence we have at least 2 solutions at least half the time. The third point is a problem when the number of solutions is odd: in that case, we may fail to project on a pair. As we have balanced superpositions, the probability of obtaining a singleton is 
half the probability of obtaining a pair. Hence, we fail to project with probability at most $1/3$.

The case of Algorithm~\ref{alg:cjs-comb} is very similar. In~\cite{DBLP:journals/jmc/ChildsJS14}, the authors chose $r = k -\log(k)$. As the output labels depend on the sum of the inputs, they can be larger, and the output space is of size $k2^r$. Hence, in~\cite{DBLP:journals/jmc/ChildsJS14}, a value of $r=k - \log(k)$ is chosen, to guarantee a constant success probability. Moreover, a step of rejection sampling (still with constant probability) is added, to guarantee a uniform sampling in the smaller interval.

To summarize, the two routines take $k$ elements, and produce with constant probability $p$ one refined element which is $r \simeq k$ bits better, at a cost of $\widetilde{O}(2^{ck})$, with $c$ the exponent of finding the solutions in the combination routines.

\subsection{CJS algorithm~\cite{DBLP:journals/jmc/ChildsJS14}}\label{orig}
If we want each routine in the pipeline to be the same, we will
have $m$ routines, and we need $mr \simeq n$ in order to succeed. The total cost is then of $(k/p)^m$ queries, $km$ qubits (excluding the quantum oracle overhead), a classical time in $\widetilde{O}(\sum_{i < m} (k/p)^i2^{k}) = \widetilde{O}((k/p)^m2^{k})$ and a polynomial classical memory.
If $ k = \alpha\sqrt{n\log_2(n)}$, the query cost is in $L(1/(2\alpha)$, and the time cost is in $L(1/(2\alpha) + \alpha$.
The classical cost is minimized for $\alpha = 1/\sqrt2$, which implies $k = \sqrt{n\log_2(n)/(2)}$, which leads to a quantum query cost in $L(1/\sqrt{2})$ and a classical time cost in $L(\sqrt{2})$.

\begin{remark}
 The quantum query exponent of~\cite[Theorem 5.2]{DBLP:journals/jmc/ChildsJS14} is not tight.
\end{remark}

\section{Subset-sum algorithms}\label{subset-sum}

It turns out that solving $x \cdot (\ell_1,\dots,\ell_k) \mod 2^r = V$ with $x \in \{0,1\}^k$ for random instances can be done more efficiently than brute-force. Indeed, as we have a small number of solutions, it corresponds to a random instance
of a subset-sum problem with a density close to 1, for which multiple algorithms have been proposed~\cite{DBLP:journals/siamcomp/SchroeppelS81,DBLP:conf/eurocrypt/BeckerCJ11,DBLP:conf/tqc/HelmM18}.
All these algorithms (classical and quantum) have an exponential complexity, in $\widetilde{O}(2^{cn})$, for a given constant $c < 1$. As we consider a polynomial-quantum memory algorithm, we will focus on the classical subset-sum algorithms.

These algorithms rely on list-merging techniques: the complete solution is contructed from lists of candidate partials solutions.

A simple example is the Schroppel-Shamir algorithm~\cite{DBLP:journals/siamcomp/SchroeppelS81}. The space of possible solutions of $\sum_{i = 1}^{n} \varepsilon_ix_i = V$ is split into 4 parts $S_1, S_2, S_3, S_4$, with $S_1 = \sum_{i\leq n/4}\varepsilon_ix_i$, and so on.

The lists contains the possible partial sum on a fourth of the variables. The intermediate lists $S_{12}$ and $S_{34}$ contains the partial sums on the first and second half of the variables.

Without any other technique, the splitting would not gain anything, as the two intermediate lists would be of size $2^{n/2}$. The Schroeppel-Shamir algorithm gains by guessing $S_{12} \mod 2^{n/4}$.
With that guess, the list is expected to be of size $2^{n/4}$. Conversely, $S_{12} \mod 2^{n/4}$ imposes the value $V - S_{12} \mod 2^{n/4}$ for $S_{34}$. Hence all the lists are expected to be of size $\widetilde{O}(2^{n/4})$. The solution will only be found for the correct guess of the intermediate value, requiring $2^{n/4}$ guesses overall, for a total cost of $\widetilde{O}(2^{n/2})$ time, but only $\widetilde{O}(2^{n/4})$ memory.

Finally, the merging in itself is the efficient generation of the intermediate lists (and of the complete solution) from the previous lists. As we want to produce a list of values with a constrained sum, we can sort the first list, and then check for each element of the second list if there is an element leading to a correct sum in the first list. The cost of the merging is the cost of sorting the input list and constructing the output list, here $\widetilde{O}(2^{n/4})$.

The algorithm from~\cite{DBLP:conf/eurocrypt/BeckerCJ11} uses similar techniques, but with a different splitting. Instead of considering a subset of the variables, they considered a partial sum with a smaller number of terms in it. They also allowed the exponent of the subset sum in the intermediate step to lie in $\{-1,0,1\}$. With this approach, the merging has also to check for the consistency of the solutions, as the variables may overlap.

By splitting the sum in 8 and carefully choosing the size of the intermediate constraint and the ratio of -1, the authors of~\cite{DBLP:conf/eurocrypt/BeckerCJ11} obtained a complexity in $\widetilde{O}(2^{0.291n})$ in classical time and memory.

\begin{figure}
 \centering
 \begin{tikzpicture}
  \draw[loop] (0,0) -- node[name=s1]{} (1,0) -- (1,2) -- node[above] {$S_1$} (0,2) --  (0,0);
  \draw[fill=gray] (0,0) rectangle (0.25,2);
  \draw[loop] (2,0) -- node[name=s2]{} (3,0) -- (3,2) --  node[above] {$S_2$} (2,2) --  (2,0);
  \draw[fill=gray] (2.25,0) rectangle (2.5,2);
  \draw[loop] (4,0) -- node[name=s3]{} (5,0) -- (5,2) -- node[above] {$S_3$} (4,2) --  (4,0);
  \draw[fill=gray] (4.5,0) rectangle (4.75,2);
  \draw[loop] (6,0) -- node[name=s4]{} (7,0) -- (7,2) --  node[above] {$S_4$} (6,2) --  (6,0);
  \draw[fill=gray] (6.75,0) rectangle (7,2);
  \draw[loop] (1,-1) -- (1,-3) --  node[name=ss12] {} (2,-3) --  (2,-1) --  node[above,name=s12] {$S_{12}$} (1,-1);
  \draw[fill=gray] (1,-1) rectangle (1.5,-3);
  \draw[->] (s1) -- (s12.north);
  \draw[->] (s2) -- (s12.north);
  \draw[loop] (5,-1) -- (5,-3) -- node[name=ss34] {}  (6,-3) --  (6,-1) --  node[above,name=s34] {$S_{34}$} (5,-1);
  \draw[fill=gray] (5.5,-1) rectangle (6,-3);
  \draw[->] (s3) -- (s34.north);
  \draw[->] (s4) -- (s34.north);
  \node[name=1234] at (3.5,-3.5) {$S$};
  \draw[->] (ss12) -- (1234.north);
  \draw[->] (ss34) -- (1234.north);
 \end{tikzpicture}
  \caption{Schroeppel-Shamir merging}\label{merge}
\end{figure}
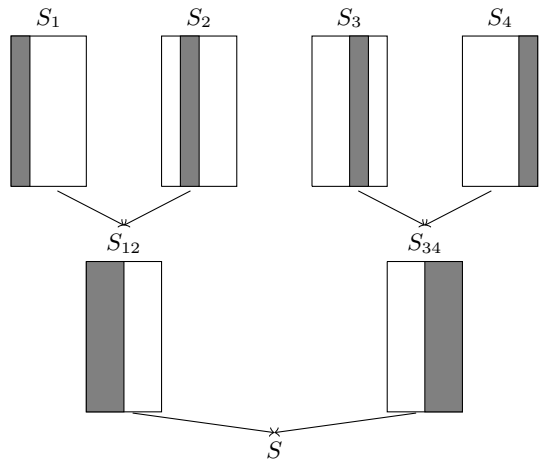

It is also possible to devise polynomial-memory algorithms for this problem. The method in~\cite{DBLP:conf/eurocrypt/BeckerCJ11} consists in merging only two lists, using a memoryless collision-finding algorithm, with a cost in $\widetilde{O}(2^{0.72n})$.

In our case, we have some instances with more than one solution. As the expected number of solutions is fixed, this does not change the asymptotic cost for constructing all the solutions.

Finally, finding the pairs for Algorithm~\ref{alg:cjs-comb} consists in finding the solutions of the slightly different equation $\lfloor x \cdot (\ell_1,\dots,\ell_k)/2^r\rfloor = V$. This equation is slightly different, but the exact same techniques can be applied, as the merging can also check for values lying in an interval. Hence, the cost is similar for both. 

For the polynomial variant, we do not have an equality to check between the two lists. This can be solved by truncating the first $r+1$ bits of the partial sum. In that case, two partial sums that lead to the correct interval have more than 50\% chance to have the same truncated value. In the case that we missed, we only need to redo this by adding $2^r$ to the partial sum before truncating. We will
also have a fixed proportion of false positives that we need to discard. Overall, this does not change the complexity cost of $\widetilde{O}(2^{cn})$.

\section{Improved hidden-shift algorithms}
\subsection{Improving the CJS algorithm}\label{improv}
Instead of using brute-force to compute the output qubit value of each routine, we can use a subset-sum algorithm, which costs $\widetilde{O}(2^{cn})$. The only difference is that the classical part cost less. With  $k = \alpha\sqrt{n\log_2(n)}$, we still have a quantum query cost in $L(1/(2\alpha))$, but the classical time cost becomes $L(1/(2\alpha) + c\alpha)$ and the optimal parameter becomes $k = \sqrt{n\log_2(n)/(2c)}$, leading to a quantum query complexity in $L(\sqrt{c/2})$ and a classical time complexity in $L(\sqrt{2c})$.

The classical time exponent of this hybrid algorithm is double the quantum query cost.

\subsection{Minimizing the classical time}\label{minclass}
The classical cost of the previous approach is almost all in the first routine of the pipeline: the second routine is called $k/p$ time less, the third $(k/p)^2$ time less, and so on. We can change the size of each routine and increase $k$ to have each routine to cost overall roughly the same. The cost of a routine is in $2^{ck}$, hence $k$ can increase by $\log_2(n)/(2c)$ for each routine. The optimal point is for $k$ increasing from 0 to $\sqrt{n\log_2(n)/c}$, by steps of $\log_2(n)/(2c)$.
With these parameters, the routine $i$ uses $k_i = i\log_2(n)/(2c)$. We need to have $\sum_{i < m} k_i \simeq n$, which implies $m \simeq 2\sqrt{cn/\log_2(n)}$. Hence, the cost of each routine will be in $2^{ck_m+o(k_m)}$. As $k_m \simeq \sqrt{n\log_2(n)/c}$, we obtain a
quantum time cost in $L(\sqrt{c})$. The first routine does nothing, hence its cost is the number of queries, which is also in $L(\sqrt{c})$.

\subsection{Enforcing a quadratic gap}\label{quad}
Section~\ref{improv} had an algorithm with $k_i = \beta\sqrt{n\log_2(n)}$, Section~\ref{minclass} used $ k_i = i\alpha\sqrt{n\log_2(n)}$. We can generalize this and consider 
It turns out the two previous approaches can be generalized, and we can consider routine input sizes of the form $ k_i = (i\alpha + \beta)\sqrt{n\log_2(n)}$, and for example ensure a quadratic gap between the number of queries and
the classical time. If we want to have each $m$ routine to cost the same, we still need $\alpha = \log_2(n)/(2c)$. This implies $ m \simeq 2c\left(-\beta+\sqrt{\beta^2+\frac1c}\right)\sqrt{n/\log_2(n)}$, hence $k_m \simeq \sqrt{\beta^2+\frac1c}\sqrt{n\log_2(n)}$.
The log of number of queries is in $c(k_m-k_1) = c\alpha m = c\left(-\beta + \sqrt{\beta^2+\frac1c}\right)\sqrt{n\log_2(n)}$, as all the steps cost the same. Enforcing a quadratic gap between the two means that $\sqrt{\beta^2+\frac1c} = 2\beta$,
hence $\beta = 1/\sqrt{3c}$. With these values, the quantum query cost is in $L\left(\sqrt{\frac{c}3}\right)$, and the classical time cost is in $L\left(2\sqrt{\frac{c}3}\right)$.

As it is similar to the classical/quantum gap of Grover's algorithm, this approach can be interesting for cases where the targeted hardness is defined with respect to some exhaustive search, as for example in the security levels 1, 3 and 5 of the NIST call for post-quantum primitives.

\subsection{Minimizing the number of queries}\label{minq}
The simplest way to minimize to number of queries is to have only one routine, and directly use Algorithm~\ref{alg:reg-comb} or Algorithm~\ref{alg:cjs-comb} to solve the whole problem: in that case, the number of queries is in $O(n)$ to obtain one target element, hence overall the quantum query cost is in $O(n^2)$. The classical cost is in $\widetilde{O}(2^{cn})$. Interestingly, this beats the asymptotic cost of the best known classical algorithm if $c < 1/2$.

\subsection{Algorithms complexity}
From Section~\ref{subset-sum}, we can take $c = 0.291$ for a memory-heavy algorithm, or $c = 0.72$ for a polynomial-memory algorithm. This leads to the exponents of Table~\ref{cost:c}.

If we only consider quantum algorithms (or that classical and quantum time and memory are equivalent), then we have sightly different complexities. 
Currently, an algorithm with the exponent $c = 0.241$ has been proposed in~\cite{DBLP:conf/pqcrypto/BernsteinJLM13}, and one with exponent $c = 0.226$ in~\cite{DBLP:conf/tqc/HelmM18}. The costs are summarized in Table~\ref{cost:q}. It is to be noted that the non-polynomial quantum memory hidden shift algorithms we obtain perform asymptotically worse than the algorithms from~\cite{DBLP:conf/asiacrypt/BonnetainN18,DBLP:journals/siamcomp/Kuperberg05,DBLP:conf/tqc/Kuperberg13}, which use a different
combination routine.

Using Grover's algorithm for a small number of queries produces an algorithm with the same cost as the exhaustive search, with a small number of queries. Overall, this is slightly worse than the approach of~\cite{DBLP:conf/stacs/EttingerH99}, which performs an exhaustive search, but achieves
a linear number of queries.

\begin{table}
\centering{}
 \begin{tabular}{|c|c|c|c|c|}
 \sline
 Quantum query & Quantum time & Quantum memory & Subset-sum & Source\\
 \sline
 $L(0.5)$ & $L(1)$ & $O(n)$ & Grover &Section~\ref{improv}\\
 \sline
 $L(1/\sqrt{2})$ & $L(1/\sqrt{2})$ & $O(n)$ & Grover &Section~\ref{minclass}\\
  \sline
 $O(n^2)$ & $2^{0.5n}$ & $O(n)$ & Grover &Section~\ref{minq}\\
  \sline
 $L(0.283)$ & $L(0.567)$ & $L(0.283)$ & \cite{DBLP:conf/pqcrypto/BernsteinJLM13} &Section~\ref{quad}\\
 \sline
 $L(0.491)$ & $L(0.491)$ & $L(0.491)$ & \cite{DBLP:conf/pqcrypto/BernsteinJLM13} &Section~\ref{minclass}\\
 \sline
 $O(n^2)$ & $\widetilde{O}\left(2^{0.241n}\right)$ & $\widetilde{O}\left(2^{0.241n}\right)$ & \cite{DBLP:conf/pqcrypto/BernsteinJLM13} &Section~\ref{minq}\\
 \sline
 $L(0.274)$ & $L(0.549)$ & $L(0.274)$ & \cite{DBLP:conf/tqc/HelmM18} &Section~\ref{improv}\\
 \sline
 $L(0.475)$ & $L(0.475)$ & $L(0.475)$ & \cite{DBLP:conf/tqc/HelmM18} &Section~\ref{minclass}\\
 \sline
 $O(n^2)$ & $\widetilde{O}\left(2^{0.226n}\right)$ & $\widetilde{O}\left(2^{0.226n}\right)$ & \cite{DBLP:conf/tqc/HelmM18} &Section~\ref{minq}\\
 \hline
\end{tabular}
\vspace{0.2cm}
\caption{Purely quantum algorithm costs}\label{cost:q}
\end{table}

\section{Conclusion}
In this paper, we showed how to use subset-sum algorithms to reduce significantly the cost of a quantum hidden shift algorithm, and proposed different quantum/classical cost tradeoffs, allowing to divide the
exponent by roughly 2.6 compared to~\cite{DBLP:journals/jmc/ChildsJS14}, and even by 4.5 for the quantum query cost if we allow a quadratic gap between the classical time cost and the quantum query cost.

\subsubsection{Improving the complexity.} In order to obtain more efficient algorithms, one might study what happens if the combined elements can be chosen from a larger pool, as done in the quantum-memory heavy algorithms of~\cite{DBLP:journals/siamcomp/Kuperberg05}. This may allow to reduce the time cost at the expense of the quantum memory, while still offering a tradeoff between classical time and quantum query. Another approach would be to apply similar techniques to the algorithm of~\cite{DBLP:conf/tqc/Kuperberg13}, if applicable. 

\bibliographystyle{splncs03}
\bibliography{biblio}

\begin{thebibliography}{10}
\providecommand{\url}[1]{\texttt{#1}}
\providecommand{\urlprefix}{URL }

\bibitem{conf/eurocrypt/AlagicR17}
Alagic, G., Russell, A.: {Quantum-Secure Symmetric-Key Cryptography Based on
  Hidden Shifts.} In: Coron, J.S., Nielsen, J.B. (eds.) {EUROCRYPT (3)}.
  {LNCS}, vol. 10212, pp. 65--93 (2017)

\bibitem{DBLP:conf/eurocrypt/BeckerCJ11}
Becker, A., Coron, J., Joux, A.: Improved generic algorithms for hard
  knapsacks. In: Paterson, K.G. (ed.) Advances in Cryptology - {EUROCRYPT} 2011
  - 30th Annual International Conference on the Theory and Applications of
  Cryptographic Techniques, Tallinn, Estonia, May 15-19, 2011. Proceedings.
  Lecture Notes in Computer Science, vol. 6632, pp. 364--385. Springer (2011)

\bibitem{DBLP:conf/pqcrypto/BernsteinJLM13}
Bernstein, D.J., Jeffery, S., Lange, T., Meurer, A.: Quantum algorithms for the
  subset-sum problem. In: Gaborit, P. (ed.) Post-Quantum Cryptography - 5th
  International Workshop, PQCrypto 2013, Limoges, France, June 4-7, 2013.
  Proceedings. Lecture Notes in Computer Science, vol. 7932, pp. 16--33.
  Springer (2013)

\bibitem{DBLP:conf/asiacrypt/BonnetainN18}
Bonnetain, X., Naya{-}Plasencia, M.: Hidden shift quantum cryptanalysis and
  implications. In: Peyrin, T., Galbraith, S.D. (eds.) Advances in Cryptology -
  {ASIACRYPT} 2018 - 24th International Conference on the Theory and
  Application of Cryptology and Information Security, Brisbane, QLD, Australia,
  December 2-6, 2018, Proceedings, Part {I}. Lecture Notes in Computer Science,
  vol. 11272, pp. 560--592. Springer (2018)

\bibitem{DBLP:journals/iacr/BonnetainS18}
Bonnetain, X., Schrottenloher, A.: Quantum security analysis of {CSIDH} and
  ordinary isogeny-based schemes. {IACR} Cryptology ePrint Archive  2018,  537
  (2018)

\bibitem{cryptoeprint:2018:383}
Castryck, W., Lange, T., Martindale, C., Panny, L., Renes, J.: Csidh: An
  efficient post-quantum commutative group action. Cryptology ePrint Archive,
  Report 2018/383 (2018), \url{https://eprint.iacr.org/2018/383}

\bibitem{DBLP:journals/jmc/ChildsJS14}
Childs, A.M., Jao, D., Soukharev, V.: Constructing elliptic curve isogenies in
  quantum subexponential time. J. Mathematical Cryptology  8(1),  1--29 (2014)

\bibitem{DBLP:conf/stacs/EttingerH99}
Ettinger, M., H{\o}yer, P.: {On Quantum Algorithms for Noncommutative Hidden
  Subgroups}. In: {{STACS} 99, 16th Annual Symposium on Theoretical Aspects of
  Computer Science, Trier, Germany, March 4-6, 1999, Proceedings}. {LNCS}, vol.
  1563, pp. 478--487. Springer (1999)

\bibitem{DBLP:journals/iacr/FeoG18}
Feo, L.D., Galbraith, S.D.: Seasign: Compact isogeny signatures from class
  group actions. {IACR} Cryptology ePrint Archive  2018,  824 (2018)

\bibitem{DBLP:conf/asiacrypt/FeoKS18}
Feo, L.D., Kieffer, J., Smith, B.: Towards practical key exchange from ordinary
  isogeny graphs. In: Peyrin, T., Galbraith, S.D. (eds.) Advances in Cryptology
  - {ASIACRYPT} 2018 - 24th International Conference on the Theory and
  Application of Cryptology and Information Security, Brisbane, QLD, Australia,
  December 2-6, 2018, Proceedings, Part {III}. Lecture Notes in Computer
  Science, vol. 11274, pp. 365--394. Springer (2018)

\bibitem{DBLP:conf/tqc/HelmM18}
Helm, A., May, A.: Subset sum quantumly in 1.17\({}^{\mbox{n}}\). In: Jeffery,
  S. (ed.) 13th Conference on the Theory of Quantum Computation, Communication
  and Cryptography, {TQC} 2018, July 16-18, 2018, Sydney, Australia. LIPIcs,
  vol. 111, pp. 5:1--5:15. Schloss Dagstuhl - Leibniz-Zentrum fuer Informatik
  (2018)

\bibitem{DBLP:journals/siamcomp/Kuperberg05}
Kuperberg, G.: {A Subexponential-Time Quantum Algorithm for the Dihedral Hidden
  Subgroup Problem}. {SIAM} J. Comput.  35(1),  170--188 (2005)

\bibitem{DBLP:conf/tqc/Kuperberg13}
Kuperberg, G.: {Another Subexponential-time Quantum Algorithm for the Dihedral
  Hidden Subgroup Problem}. In: Severini, S., Brand{\~a}o, F.G.S.L. (eds.) {8th
  Conference on the Theory of Quantum Computation, Communication and
  Cryptography, {TQC} 2013, May 21-23, 2013, Guelph, Canada}. {LIPIcs},
  vol.~22, pp. 20--34. Schloss Dagstuhl - Leibniz-Zentrum fuer Informatik
  (2013)

\bibitem{quant-ph/0406151v1}
Regev, O.: {A Subexponential Time Algorithm for the Dihedral Hidden Subgroup
  Problem with Polynomial Space}. CoRR  (2004)

\bibitem{DBLP:journals/siamcomp/SchroeppelS81}
Schroeppel, R., Shamir, A.: A t=o(2\({}^{\mbox{n/2}}\)),
  s=o(2\({}^{\mbox{n/4}}\)) algorithm for certain np-complete problems. {SIAM}
  J. Comput.  10(3),  456--464 (1981)

\bibitem{DBLP:journals/siamcomp/Shor97}
Shor, P.W.: Polynomial-time algorithms for prime factorization and discrete
  logarithms on a quantum computer. {SIAM} J. Comput.  26(5),  1484--1509
  (1997)

\end{thebibliography}

\end{document}